# Robustness of the Closest Unstable Equilibrium Point Along a P-V Curve

R. Owusu-Mireku, *Student Member, IEEE*, and H.-D. Chiang, *Fellow, IEEE*

*Abstract*—In this paper, we study numerically the behavior of the closest unstable equilibrium point (UEP) on the stability boundary of a stable equilibrium point (SEP) of a post-switching power system along a P-V curve. Using the structure-preserving model of the WSCC 9-bus 3-machine system, we show that along the load curve, the closest UEP can switch to a new UEP. We also show that the stability region of the post-switching SEP can expand and contract as the load moves towards the nose point of the P-V curve. Our numerical results also show the impact of the direction of movement of the closest UEP on the size and shape of the stability region of a SEP.

## I. INTRODUCTION

ONE approach to the transient stability analysis and control of a power system is to use direct or energy-based methods [1-3]. These methods are based on the Lyapunov function theory [1]. Direct methods assess the transient stability of a power system event by first approximating the stability region of the stable equilibrium point (SEP) for the post-event system as an energy set with an upper bound represented by a constant energy level. The constant energy level is defined by the energy at a critical point on the stability boundary of the SEP. Direct methods then check if the energy at the post-event initial state is in the energy set or not. If the energy at the post-event initial state is in the energy set then the initial state is in the estimated stability region; hence, the post-event system is stable, otherwise it might be unstable. The energy is calculated using a Lyapunov function or an energy function [1, 9]. A lot of methods have been proposed for the determination of the critical point [1-8].

One of these methods is the closest unstable equilibrium point (UEP) method. The closest UEP method uses the UEP with the lowest Lyapunov or energy function value on the stability boundary of a post-event SEP as the critical point. The concept of the closest UEP has been used to estimate the stability region of a power system since the early 1970s [4, 5]. The computation of the closest UEP is currently a very challenging task, despite the large amount of research and effort that has been dedicated towards the efficient and reliable computation of the closest UEP over the years [6-8].

Due to the challenges associated with computing the closest UEP, it is important to study its robustness to changes in the power grid. This is because if the closest UEP is robust to a parameter change in the power grid, then once it is computed for one value of the parameter, the closest UEP for other values of the parameter could be traced from it. These parameter changes could be changes in network topology, dynamic parameter changes, or changes in the loading condition of the power system for a defined stress pattern. The robustness of the closest UEP to a loading stress pattern is particularly important in applications where the closest UEP is used for transient stability constrained available transfer capability calculations [10]. Since a robust closest UEP along a stress pattern will imply that once the closest UEP is computed at the base-case loading condition, it can be easily computed for all other loading conditions along a load/P-V curve. A closest UEP that is robust along a load curve will also imply that the closest UEP for a base-case loading condition could be traced or obtained from the load curve of the post-event system, since we know that it is the closest UEP that coincides with the SEP at the saddle node bifurcation [11].

In [1] the author presents and proves the invariant property of the closest UEP with respect to machine inertia and damping. The author also presents a theory on the robustness of the closest UEP to network topology changes and changes in real power injection. In [12] the changes in UEPs on the stability boundary of a SEP for a reduced power system model under varying loading conditions was studied. The authors showed that, under heavy loading conditions, a UEP on a stability boundary can disappear.

In this work, we extend the work in [1, 12] by numerically exploring the robustness of the closest UEP with respect to changes in loading conditions for a given stress pattern. We demonstrate that along a load/P-V curve of a power system, there can be a bifurcation such that the closest UEP on the stability boundary of a SEP for a given energy function can change to a new UEP, with the old closest UEP jumping off the boundary of the SEP and the new closest UEP jumping onto the boundary of SEP in state space. We also show with numerical examples that there can be a mismatch between changes in the size of the machine angle stability region and the voltage margin along the P-V curve for a given stress pattern. Thus, we show that in some cases, while moving along the P-V curve towards the nose point, the angular stability region expands, even though the voltage margin is obviously decreasing. Our numerical simulation also shows that for a SEP whose stability boundary is unbounded and has only one UEP on its stability boundary, the expansion or contraction of the stability region of the SEP along the P-V curve depends on the direction of motion

Robert Owusu-Mireku and Hsiao-Dong Chiang are with the School of Electrical and Computer Engineering, Cornell University, Ithaca, NY 14853, USA (e-mail: ro82@cornell.edu, hc63@cornell.edu).

of the closest UEP on its boundary with respect to the SEP; as the closest UEP moves closer to the SEP, the stability boundary contracts, and as it moves away from the SEP, the stability boundary expands.

In Section II, we provide mathematical definitions. We then present the problem definition and methodology in Section III. Section IV presents our numerical case study with the structure-preserving model of the WSCC 9-bus 3-machine system. Finally, we discuss our concluding remarks in Section V.

## II. Definitions

### A. Stability Region Definitions

The stability region or region of attraction $A(x_s)$ of an asymptotically stable equilibrium point (SEP) $x_s$ of an ordinary differential equation $\dot{x} = f(x)$ is defined as:

$$A(x_s) := \left\{ x \in R^n : \lim_{t \to \infty} \varphi(t, x) = x_s \right\}. \tag{1}$$

For a SEP $x_s$, if all the equilibrium points on its stability boundary are hyperbolic, the stable and unstable manifolds of the equilibrium points satisfy the transversality condition [1], and every trajectory on the stability boundary converges to an equilibrium point as $t \to \infty$, then the stability boundary $\partial A(x_s)$ of $x_s$ is defined as the union of the stable manifolds $W^s(x_i)$ of the unstable equilibrium points (UEP) $x_i$ on $\partial A(x_s)$ where $i = 1, 2 \cdots n$ and $n$ is the number of unstable equilibrium points on $\partial A(x_s)$ [1].

### B. Dynamic Model of a Power System

The power system dynamic model can be represented by a system of differential algebraic equations (DAE), as shown in (2) where the differential equation represents the electrical and electromechanical dynamics of generators, their controls, dynamic loads, and other dynamically modeled components. The algebraic equations represent the network topology and other static relationships in the system.

$$\begin{aligned} \dot{x} &= f(x, y) \\ 0 &= g(x, y) \end{aligned} \tag{2}$$

The general form of the DAE model in (2) has been thoroughly analyzed in [1]. The stability boundary of (3) has also been characterized in [1] as comprising two parts: the union of the stable manifolds of an unstable equilibrium point on the stability boundary, and a collection of trajectories reaching singular surfaces. For the application of the energy-based direct methods to systems of the form (2), (2) can be approximated by using the singular perturbation approach (SPA) [1].

In the singular perturbation approach, we replace (2) with a two-time scale system with a slow variable $x$ and a fast variable $y$ of the form:

$$\dot{x} = f(x, y), \qquad \varepsilon \dot{y} = g(x, y) \quad t > t_s \tag{3}$$

where $\varepsilon$ is a sufficiently small positive scalar. The author in [1] has shown that we can use the stability boundary and region of the singular perturbed system (3) as an approximation of the stability boundary and region of the DAE system (2).

### C. Energy Function

A function $V: R^n \to R$, is an energy function for an ODE, say (3), if the following three conditions are satisfied:
1. Along any nontrivial trajectory $\varphi(t, x_0, y_0)$, $\dot{V}(\varphi(t, x_0, y_0)) \leq 0$ and the set $\{t \in R : \dot{V}(\varphi(t, x_0, y_0)) = 0\}$ has a measure zero in R.
2. If $\{V(\varphi(t, x_0, y_0)) : t \geq 0\}$ is bounded, then $\{\varphi(t, x_0, y_0) : t \geq 0\}$ is bounded.

### D. The Closest UEP

A UEP $x_p$ on the stability boundary of a SEP $x_s$ is the closest UEP of the SEP with respect to an energy function $V(.)$,

$$\text{if } V(x_p) = \min_{x_i \in \partial A(x_s)} V(x_i).$$

The closest UEP of $x_s$ of a system, say (3), with respect to the system's energy function $V(.)$ exists and is unique. The closest UEP $x_p$ is important in estimating the stability region of the SEP $x_s$ because $x_p$ is the point of minimum energy on the stability boundary of $x_s$. Thus, the stability region characterized by the constant energy level defined by the closest UEP $x_p$ is the largest energy level that is entirely contained within the stability region of $x_s$; hence, it is optimal.

These characteristics of the closest UEP makes it very important for the estimation of the stability region of a SEP, like a power system's stable equilibrium state. However, given a SEP $x_s$, it is very difficult to find its closest UEP. Consequently, the robustness of a computed closest UEP to changes in network topology, dynamic parameters, or the behavior of the closest UEP along the P-V curve for a given load stress pattern of a power systems is very important in the dynamic stability analysis of power systems. In this work, we focus on the latter.

## III. Problem Definition and Methodology

### A. Problem Definition

In this work, we studied the behavior of the closest UEP and consequently the stability region of post-switching SEPs along a P-V curve for a given stress pattern. The changes in load and generation were made in the pre-switching system. Thus, the generic model for the system used in this study can be represented as follows:

Pre-switching system:
$$\begin{aligned} 0 &= g_b(y, \lambda) = g_{pre}(y) + \lambda b \\ \dot{x} &= f_{pre}(x, y) \end{aligned} \tag{4}$$

Post-switching system:
$$\begin{aligned} \dot{x} &= f_{post}(x, y) \\ 0 &= g_{post}(x, y) \end{aligned} \tag{5}$$

where $f_{pre}(.)$ represents the dynamic component of the equilibrium equations of the pre-fault system, $g_b(.)$ represents the active and reactive power flow balance equations of the pre-switching system, $f_{post}(.)$ represents the vector field of the dynamic component of the post-switching system, $g_{post}(.)$

represents the power balance equation of the post-switching system, $b$ is the power injection variation vector for the stress pattern, and $\lambda$ is the scaling factor.

### B. Methodology

The study is implemented as follows:
1. For a given switching event and stress pattern $b$, select the loading and generation condition by selecting the scaling factor $\lambda$.
2. Run power flow for the pre-switching system.
3. Initialize the dynamic variables, $x$, for the loading condition.
4. Using the singular perturbed system (SPS):
   a. Solve for the post-switching equilibrium point, with a Newton-based method and confirm with a time domain simulation.
   b. Construct the stability region of the post-switching SEP.
   c. Compute the equilibrium points in the neighborhood of the stability region of the post-switching SEP using the method proposed in [14].
   d. Assess the type of equilibrium points and then plot the UEPs that are either on the stability boundary or in the neighborhood of the constructed stability region.
   e. Find the closest UEP on the stability boundary of the post-switching SEP using the energy function in [9].

Given a switching event, these steps are implemented for several loading conditions along the P-V curve until there is a structure induced or saddle node bifurcation in the pre-switching system or in the post-switching system.

The stability region of the post-switching SEP is constructed by creating a grid of initial points around the UEP in the machine angle space. The solutions of the equilibrium equations for the dynamic system, starting at these initial points, are then computed using time domain simulation. If the L2 norm of the difference between the computed equilibrium point and the SEP is below a defined threshold, then the initial point is in the stability region of the SEP for that algebraic solver. Since the stability region must be connected, any small unconnected portion can be discarded, as they are usually the result of a numerical approximation or ill-conditioned Jacobian. Use a different ODE solver if the unconnected portion is large.

## IV. NUMERICAL STUDY

The study is performed on the structure-preserving model of the WSCC 9-bus 3-machine system with classical generators and a constant impedance load model. See [1] for the system data. The generalized list of equations for the structure-preserving model in the center of inertia (COI) reference frame is as shown below.

For $n$ generators and $m$ buses,
$$\dot{\tilde{\delta}}_i = \tilde{\omega}_i \quad (6)$$
$$M_i \dot{\tilde{\omega}}_i = -D_i \tilde{\omega}_i + P_{m_i} - \frac{E'_{qi} V_i \sin(\tilde{\delta}_i - \tilde{\theta}_i)}{X'_{di}} - \frac{M_i}{M_i} P_{COI} \quad (7)$$

For generator buses $i = 1, \ldots, n$:
$$\left(I_{di} + jI_{qi}\right) e^{-j(\delta_i - \pi/2)} = \sum_{k=1}^{m} Y_{ik} e^{j\alpha_{ik}} V_k e^{j\tilde{\theta}_k}$$
$$I_{di} = \frac{E'_{qi} - V_i \cos(\tilde{\delta}_i - \tilde{\theta}_i)}{X'_{di}}, \quad I_{qi} = \frac{V_i \sin(\tilde{\delta}_i - \tilde{\theta}_i)}{X'_{qi}} \quad (8)$$

For load buses $i = n+1, \ldots, m$:
$$0 = \sum_{k=1}^{m} Y_{ik} e^{j\alpha_{ik}} V_k e^{j\tilde{\theta}_k} \quad (9)$$

$$\delta_0 = \frac{1}{M_T} \sum_{i=1}^{n} M_i \delta_i, \quad \omega_0 = \frac{1}{M_T} \sum_{i=1}^{n} M_i \omega_i$$
$$M_T = \sum_{i=1}^{n} M_i, \tilde{\delta}_i = \delta_i - \delta_0, \tilde{\omega}_i = \omega_i - \omega_0,$$
$$\tilde{\theta}_i = \theta_i - \theta_0 \text{ for } i = 1, \ldots, n,$$
$$P_{COI} = \sum_{i=1}^{n} P_{m_i} - \sum_{i=1}^{n} \frac{E'_{qi} V_i \sin(\tilde{\delta}_i - \tilde{\theta}_i)}{X'_{di}}$$

where $\delta_i, \omega_i, M_i, D_i, P_{m_i}, E'_{qi}, X'_{di}, V_i, \theta_i$, and $Y_{ik} e^{j\alpha_{ik}}$ are the rotor angle of machine $i$, speed of machine $i$, moment of inertia of machine $i$, the damping of machine $i$, mechanical power of machine $i$, equivalent transient quadrature internal voltage of machine $i$, equivalent direct transient reactance of machine $i$, the voltage magnitude at bus $i$, the voltage angle at bus $i$, and the network admittance between buses $i$ and $k$, respectively.

The singular perturbed system for the above DAE with $\varepsilon$ of 0.001 is used for the simulation and construction of the stability region. A hybrid integration technique combining the trapezoidal method and the implicit Euler method with an adaptable step size is used in the computation of the stability regions. The convergence criteria used for the stability region construction is an L2 norm of 0.1. The stress pattern used for the load curve is a proportional increase in load at all buses at a constant power factor with the change in load supported by only the slack bus. The study is performed by observing the changes in the dynamics of the post-switching system for different loading conditions along the P-V curve for the stress pattern above. The switching events used in the study are the same as those used for the numerical simulation of the WSCC 9-bus 3-machine system in [13].

### A. Behavior of the Closest UEP Along the P-V Curve

In this section, we study the behavior of the closest UEP on the stability boundary of the post-switching SEP along a P-V curve for the WSCC 9-bus 3-machine system after opening the line between buses 4 and 5. We focus on the behavior of two UEPs, UEP1 and UEP2, in the neighborhood of the post-switching SEP. Fig. 1(a) shows the projection of the stability region of a post-switching SEP into the $\delta_1 - \delta_2$ machine angle space for the base-case loading condition. The green region represents the stability region of the SEP, the blue star represents the closest UEP (called UEP1), and the red star represents the only other UEP in the neighborhood of the post-switching SEP, called UEP2. There is only one UEP, UEP1, on the stability boundary of this SEP, implying the stability boundary of this SEP is unbounded and partially defined by the stable manifold of UEP1 [1]. UEP1 is the closest UEP for this post-switching SEP, since it is the only UEP on the stability boundary of the post-switching SEP, as shown in Fig. 1(a). From Fig. 1(b), we observe that when the loading condition moves along the P-V curve to 1.4 ($\lambda=0.4$) of the base load, a bifurcation occurs where UEP1 moves off the stability boundary of the post-switching SEP while UEP2 moves onto the stability boundary. Thus, at this loading condition, UEP2 is the only UEP on the stability boundary of the post-switching SEP, and hence, it is the closest UEP for the post-switching SEP. Fig. 1(c)-(d) show that as we keep moving along the P-V curve towards the nose point, UEP2 stays on the stability boundary of the posts-switching SEP while UEP1 moves away from the stability boundary. The movement of the UEPs along the P-V curve towards the nose point is also shown in Fig. 2. In Fig. 2, the solid line plots represent the path of the machine angle variables for the SEP, the long dashed line plots represent the path of the machine angles for UEP1, and the short dashed lines represent the path of the machine angles for UEP2. All orange plots are $\delta_1$ values, the red plots are $\delta_2$ values, and the blue plots are $\delta_3$ values of their respective states (SEP, UEP1, and UEP2). Thus, the orange solid line plot represents the path of $\delta_1$ for the SEP, and the red short dashed line plot represents the path of $\delta_2$ for UEP2 and so on. Fig. 2 shows that the machine angle values for UEP1 move away from the machine angle values of the SEP while the machine angle values of UEP2 move towards the values of the SEP along the P-V curve. Similar trends are also observed in the post-switching system of switching events involving line 4-6 and 7-5. See Fig. 4 for line 4–6. These observations show that the closest UEP always exists, and that there exists a loading condition $r$ beyond which the closest UEP cannot be traced from the closest UEP for the base loading condition.

### B. Changes in the Stability Region along the P-V Curve

Fig. 1(a)-1(d) also shows that the size of the stability region of the post-switching system along the P-V curve, moving from the base case towards the nose point, enlarges for a while and then starts to shrink. The number of initial points, the green asterisk, that form the stability region of the post-switching SEP changes from 973 points at the base case to 1320 points at 1.4 of the base loading condition and then reduces to 1074 at 1.6 of the base loading condition. Thus, somewhere in between 1.4 and 1.6 of the base loading condition, a bifurcation occurs where the stability region begins to shrink along the P-V curve as the loading condition moves towards the nose point. This reduction in the size of the stability region of the post-switching SEP continues until there is a structure-induced bifurcation at a loading condition of 2.39 of the base case. Similar trends are also observed in the post-switching system of switching events involving line 7-5 and 4-6. Fig. 1 also shows that the expansion of the stability region occurred when the distance between the closest UEP and the SEP was increasing, and the stability region contracted when the distance between the closest UEP and the SEP was decreasing. The relationship between the change in the closest UEP direction of movement with respect to the SEP and the change in the size of the stability region of the SEP suggests that the direction of the movement of the closest UEP determines the size of the stability region of these SEPs. Fig. 3 shows the P-V curves for the bus voltages of the three generator buses for the post-switching system. We observe that UEP1 and UEP2 have the same voltage magnitudes. This is true for all the other buses in the system. Figs. 1, 2, and 3 show that while the voltage margin is constantly decreasing along the P-V curve towards the nose point, the generator angle stability region does not necessarily decrease as you move along the P-V curve towards the nose point.

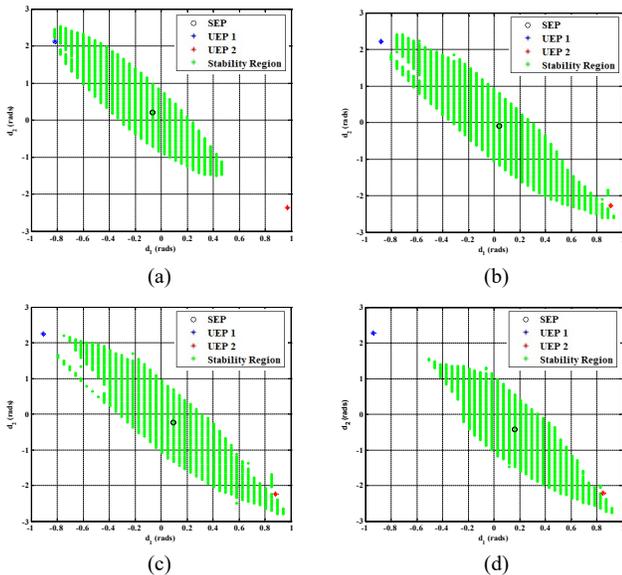

Fig. 1. Plot showing changes in the location of the closest UEP and changes in the stability region along a P-V curve for the post-switching system after opening line 4-5. (a) $\lambda = 0$. (b) $\lambda = 0.4$. (c) $\lambda = 0.6$. (d) $\lambda = 0.8444$.

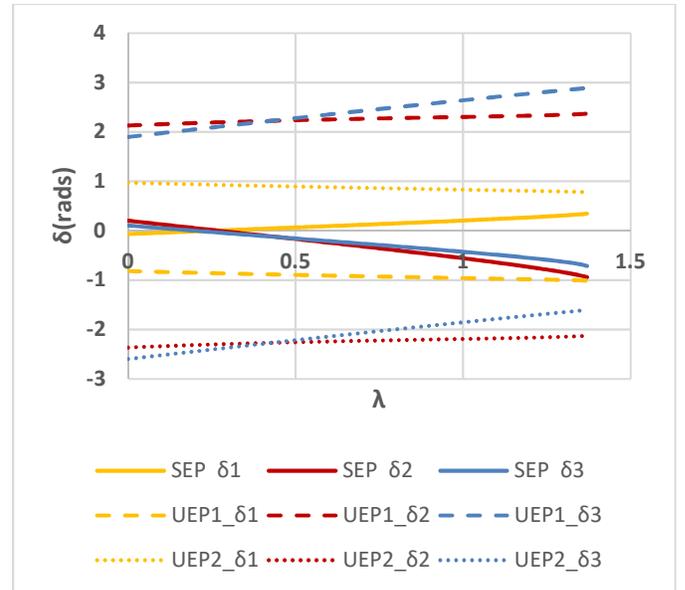

Fig. 2. Plot showing movement of the UEPs and SEP along the P-V curve towards the nose point.

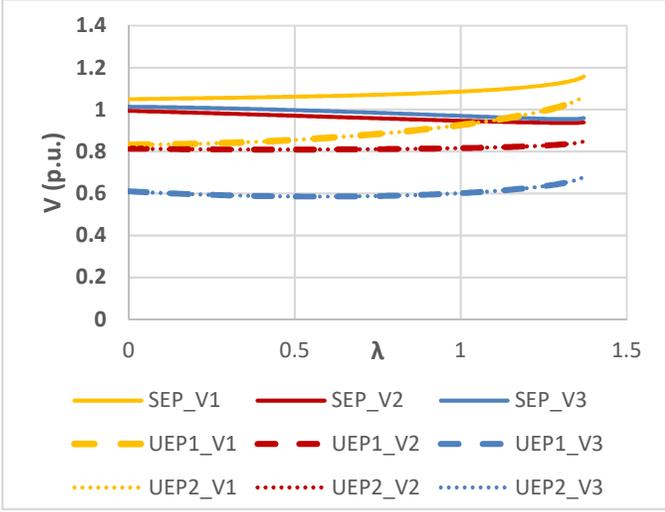

Fig. 3. Post-switching P-V curve showing the path of the bus voltages for the three generator buses.

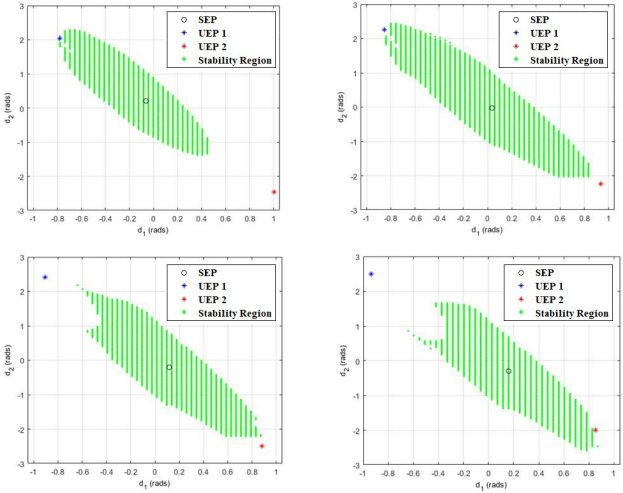

Fig. 4. Plot showing changes in the location of the closest UEP and changes in stability region along a P-V curve for the post-switching system after opening line 4 - 6. (a) $\lambda = 0$. (b) $\lambda = 0.4$. (c) $\lambda = 0.7$. (d) $\lambda = 0.8444$.

## C. Discussion

The numerical results show that the closest UEP is not fully robust along the P-V curve. Thus, along the P-V curve there might exist a certain loading condition beyond which the closest UEP cannot be predicted using the base-case closest UEP. Thus, any method that intends to calculate the closest UEP for one loading condition based on the closest UEP for another loading condition must check first to ensure the bifurcation observed above does not occur. The results also show that an increase in loading conditions does not necessarily imply a decrease in the degree of transient stability. Since, as we observed, the stability region of the post-event SEP can expand in some cases under increasing loading conditions. Does this imply we can improve stability by increasing load? We will have to find out in future work.

## V. Conclusion

In this work, we have explored the robustness of the closest UEP with respect to changes in loading conditions along a P-V curve for a given stress pattern. We have demonstrated that along the P-V curve of a power system, the closest UEP on the stability boundary of a SEP for a given energy function can change to a new UEP, with the old closest UEP leaving the stability boundary of the SEP and the new closest UEP jumping onto the stability boundary of the SEP. We have also shown that in some cases, while moving along the P-V curve towards the nose point, the angular stability region expands, even though the load margin is decreasing. Finally, we have shown numerically that for a SEP whose stability boundary is unbounded and has only one UEP on its stability boundary, the expansion or contraction of the stability region of a SEP along the P-V curve depends on the direction of motion of the closest UEP on its boundary with respect to the SEP. The application of this knowledge in the computation of closest UEPs and the computation of the transient stability constrained available transfer capability will be explored in future work.